\documentclass[twocolumn]{aastex631}
\usepackage{booktabs}
\usepackage{hyperref}
\usepackage{orcidlink}
\usepackage{amsmath}
\usepackage{natbib}
\usepackage{graphicx}
\usepackage{xcolor}

\shorttitle{Sutlej Stream in DELVE DR3}
\shortauthors{Burdett et al.}

\begin{document}
\title{A Tributary of Tucana III? The Sutlej Stellar Stream in DELVE DR3}
\newcommand{\hideemail}[1]{}

\author[0009-0006-9972-6937]{Elliott S. Burdett}
\affiliation{Department of Astronomy and DiRAC Institute, University of Washington, Seattle, WA 98195, USA}
\correspondingauthor{Elliott S. Burdett}
\email{esb30@uw.edu}

\author[0000-0003-2497-091X]{Nora Shipp}
\affiliation{Department of Astronomy and DiRAC Institute, University of Washington, Seattle, WA 98195, USA}

\author[0000-0001-6957-1627]{Peter S. Ferguson}
\affiliation{Department of Astronomy and DiRAC Institute, University of Washington, Seattle, WA 98195, USA}

\author[0000-0002-8354-7356]{Arpit Arora}
\affiliation{Department of Astronomy and DiRAC Institute, University of Washington, Seattle, WA 98195, USA}

\author[0000-0002-3475-7648]{Gourav Khullar}
\affiliation{Department of Astronomy and DiRAC Institute, University of Washington, Seattle, WA 98195, USA}
\affiliation{eScience Institute, University of Washington, Physics-Astronomy Building, Box 351580, Seattle, WA 98195-1700, USA}

\author[0000-0003-0478-0473]{Chin Yi Tan}
\affiliation{Department of Astronomy and DiRAC Institute, University of Washington, Seattle, WA 98195, USA}

\author[0000-0002-3936-9628]{Jeffrey L. Carlin}
\affiliation{NSF-DOE Vera C. Rubin Observatory / NSF NOIRLab, 950 N. Cherry Ave., Tucson, AZ 85719, USA}

\author[0000-0003-1697-7062]{William Cerny}
\affiliation{Department of Astronomy, Yale University, New Haven, CT 06520, USA}

\author[0000-0002-1763-4128]{Denija Crnojevi\'c}
\affiliation{Department of Physics \& Astronomy, University of Tampa, 401 West Kennedy Boulevard, Tampa, FL 33606, USA}

\author[0000-0001-8251-933X]{Alex Drlica-Wagner}
\affiliation{Fermi National Accelerator Laboratory, P.O.\ Box 500, Batavia, IL 60510, USA}
\affiliation{Kavli Institute for Cosmological Physics, University of Chicago, Chicago, IL 60637, USA}
\affiliation{Department of Astronomy and Astrophysics, University of Chicago, Chicago, IL 60637, USA}
\affiliation{NSF-Simons AI Institute for the Sky (SkAI), 172 E. Chestnut St., Chicago, IL 60611, USA}

\author[0000-0002-8448-5505]{Denis Erkal}
\affiliation{School of Mathematics and Physics, University of Surrey, Guildford GU2 7XH, UK}

\author[0000-0002-4863-8842]{Alexander P. Ji}
\affiliation{Department of Astronomy and Astrophysics, University of Chicago, Chicago, IL 60637, USA}
\affiliation{Kavli Institute for Cosmological Physics, University of Chicago, Chicago, IL 60637, USA}

\author[0000-0002-9269-8287]{Guilherme Limberg}
\affiliation{Kavli Institute for Cosmological Physics, University of Chicago, Chicago, IL 60637, USA}

\author[0000-0002-8093-7471]{Pol Massana}
\affiliation{NSF NOIRLab, Casilla 603, La Serena, Chile}

\author[0000-0002-8282-469X]{Noelia E. D. No\"el}
\affiliation{School of Mathematics and Physics, University of Surrey, Guildford GU2 7XH, UK}

\author[0000-0002-6021-8760]{Andrew B. Pace}
\affiliation{Department of Astronomy, University of Virginia, 530 McCormick Road, Charlottesville, VA 22904, USA}

\author[0000-0001-5805-5766]{Alexander H. Riley}
\affiliation{Lund Observatory, Division of Astrophysics, Department of Physics, Lund University, SE-221 00 Lund, Sweden}

\author[0000-0003-4102-380X]{David J. Sand}
\affiliation{Department of Astronomy/Steward Observatory, 933 North Cherry Avenue, Room N204, Tucson, AZ 85721-0065, USA}

\author[0000-0003-1479-3059]{Guy S. Stringfellow}
\affiliation{University of Colorado Boulder, Boulder, CO 80309, USA}

\author[0000-0002-3690-105X]{Julio A. Carballo-Bello}
\affiliation{Instituto de Alta Investigaci\'on, Universidad de Tarapac\'a, Casilla 7D, Arica, Chile}

\author[0000-0002-7155-679X]{Anirudh Chiti}
\affiliation{Kavli Institute for Particle Astrophysics \& Cosmology, P.O.\ Box 2450, Stanford University, Stanford, CA 94305, USA}

\author[0000-0003-1680-1884]{Yumi Choi}
\affiliation{NSF-DOE Vera C. Rubin Observatory / NSF NOIRLab, 950 N. Cherry Ave., Tucson, AZ 85719, USA}

\author[0000-0001-5160-4486]{David J. James}
\affiliation{ASTRAVEO LLC, PO Box 1668, Gloucester, MA 01931, USA}
\affiliation{Applied Materials Inc., 35 Dory Road, Gloucester, MA 01930, USA}

\author[0000-0002-9144-7726]{Clara E. Mart\'inez-V\'azquez}
\affiliation{NSF NOIRLab, 670 N. A'ohoku Place, Hilo, Hawai'i, 96720, USA}

\author[0000-0003-0105-9576]{Gustavo E. Medina}
\affiliation{Kapteyn Astronomical Institute, University of Groningen, P.O.\ Box 880, 9700 AV Groningen, The Netherlands}
\affiliation{Department of Astronomy \& Astrophysics, University of Toronto, Toronto, ON M5S 3H4, Canada}

\author[0000-0001-9649-4815]{Bur\c{c}in Mutlu-Pakdil}
\affiliation{Department of Physics and Astronomy, Dartmouth College, Hanover, NH 03755, USA}

\author[0000-0001-9438-5228]{Mahdieh Navabi}
\affiliation{School of Mathematics and Physics, University of Surrey, Guildford GU2 7XH, UK}

\author[0000-0002-1793-3689]{David L. Nidever}
\affiliation{Department of Physics, Montana State University, P.O.\ Box 173840, Bozeman, MT 59717-3840, USA}

\author[0000-0003-4341-6172]{A. Katherina Vivas}
\affiliation{Cerro Tololo Inter-American Observatory/NSF NOIRLab, Casilla 603, La Serena, Chile}

\author[0000-0001-6455-9135]{Alfredo Zenteno}
\affiliation{Cerro Tololo Inter-American Observatory/NSF NOIRLab, Casilla 603, La Serena, Chile}

\collaboration{100}{DELVE Collaboration}

% patches to remove () around collaboration
\makeatletter
\def\@collaboration@present#1#2#3#4{%
 \par
 \begingroup
  \frontmatter@collaboration@above
  \@affilID@def{}%
  \@tempcnta\z@
  \@author@present{}{\ignorespaces#3\unskip}{#4}%
  \par
 \endgroup
 \set@listcomma@list#1%
}%
\makeatother

\begin{abstract}
We present the first analysis of the Sutlej stellar stream using deep DECam photometry from the third data release of the DECam Local Volume Exploration Survey (DELVE DR3). Sutlej was discovered in \textit{Gaia} DR3 as two parallel branches of old, metal-poor stars in projection near the SMC. Our DELVE DR3 sample reaches $\sim$2.5--3.5\,mag below the main-sequence turnoff of Sutlej, and with a main-sequence matched filter built from an isochrone fit to Sutlej, we recover both previously identified branches, and find additional structure between them and at the periphery of the stream, occupying a region of $\sim$11$^\circ\times$7$^\circ$ scattered at $\sim$23--33\,kpc. We investigate this region by separating it into four segments that have Sutlej-like properties. The isochrone is also consistent with the nearby ultra-faint satellite Tucana III (Tuc III), and the two systems have similar distances and proper motions. $N$-body simulations of a Tuc III-like satellite show that, once the passage of a massive LMC is accounted for, Tuc III's preceding pericentric passage was deep enough to strip most of the progenitor. The encounter with the LMC then spreads the debris across the Sutlej field at the observed distances and proper motions, though the models do not recreate the observed discrete overdensities. We find that Sutlej is compatible with a Tuc III origin.
\end{abstract}

\section{Introduction}
\label{sec:intro}
Few tracers map the Milky Way's dark matter as directly as stellar streams, the tidal debris stripped from dwarf galaxies and globular clusters \citep[e.g.,][]{Bonaca2025}. Streams trace the Galactic potential along their orbits, and can be used to constrain the mass and shape of the dark matter halo \citep{Johnston1999, Koposov2010, LawMajewski2010, Gibbons2014, Bowden2015, Bovy2016, BonacaHogg2018}. Many streams are thin and dynamically cold, and are thus sensitive to small-scale perturbations. A close passage by a dark matter subhalo can leave a gap, create a spur, or modulate the density along a stream \citep{Ibata2002, Johnston2002, Carlberg2012, ErkalBelokurov2015, Erkal2016, Bonaca2019}, which makes streams one of the few probes of low-mass subhalos at, or below, the threshold of galaxy formation. Streams that pass near the LMC show indications of past interactions in misalignment between their tracks and proper motions \citep{Erkal2019, Shipp2019}, and these offsets have been used to measure the mass of the LMC \citep{Erkal2019, Vasiliev2021, Shipp2021}.

Stream discovery has accelerated with each generation of wide-field imaging. Isochrone filtering of photometry from the Sloan Digital Sky Survey \citep[SDSS;][]{York2000}, the VLT Survey Telescope ATLAS \citep[VST ATLAS;][]{Shanks2015}, the Panoramic Survey Telescope and Rapid Response System \citep[Pan-STARRS;][]{Chambers2016}, and the Dark Energy Survey \citep[DES;][]{DES2016} revealed dozens of streams \citep[e.g.,][]{Odenkirchen2001, Belokurov2006, Grillmair2006, Koposov2014, Bernard2016, Shipp2018}. \textit{Gaia} astrometry \citep{GaiaDR2} has made it possible to discover streams in phase space \citep{Malhan2018, Malhan2018b, Ibata2021} and to measure proper motions that separate faint members from the surrounding stellar halo \citep{PriceWhelan2018, Koposov2019, Shipp2019}. Spectroscopic surveys such as the Southern Stellar Stream Spectroscopic Survey \citep[S$^5$;][]{Li2019, Li2022, Limberg2026} and the Dark Energy Spectroscopic Instrument \citep[DESI;][]{Aganze2025, Mohammed2026} have measured radial velocities and metallicities for many streams. Nearly 100 streams are now known \citep{Mateu2023, Bonaca2025}, but many are traced only by the bright giants that \textit{Gaia} can measure, and their full extent remains poorly constrained.

The Sutlej stream was discovered by \citet[][hereafter N24]{Nidever2024} in \textit{Gaia} DR3 \citep{GaiaDR3}. N24 combined \textit{Gaia} proper motions with metallicities, effective temperatures, and surface gravities derived by \citet{Andrae2023} from the \textit{Gaia} XP spectra of 175 million stars \citep{DeAngeli2023}. In the vicinity of the SMC, N24 identified an old, metal-poor population with coherent proper motion. N24 characterized Sutlej as two nearly parallel branches, each $\sim$8$^\circ \times 0.6^\circ$, separated by $\sim$2.5$^\circ$, with a total stellar mass of $\sim$33{,}000\,M$_\odot$ and a distance of $\sim$32\,kpc. Because \textit{Gaia} reaches only $G \approx 21$, at 32\,kpc N24 sampled only Sutlej's red giant branch and horizontal branch. With deeper photometry that reaches below the main-sequence turnoff, we can map Sutlej's structure and extent with many more stars and approximate its distance from the main sequence.

Sutlej resides in a region crowded with old, metal-poor substructure. The ultra-faint satellites Tucana III (Tuc III) and Tucana IV (Tuc IV) lie within $\sim$5$^\circ$ of the N24 footprint \citep{DrlicaWagner2015}, and the Indus stream was identified nearby in DES \citep{Shipp2018}. The SMC Northern Overdensity (SMCNOD), a $\sim$6$^\circ \times 2^\circ$ structure $\sim$8$^\circ$ north of the SMC also found in DES \citep{Pieres2017}, lies at the distance of the SMC ($\sim$62\,kpc) and contains both intermediate-age and older, more metal-poor stars, most likely stripped from the SMC by the LMC \citep{Massana2024}. Tuc III in particular is at a comparable distance to Sutlej ($\sim$25\,kpc; \citealt{DrlicaWagner2015}) and is itself being tidally disrupted, which raises the possibility that the two are related. Tuc III was discovered in the second year of DES data \citep{DrlicaWagner2015} with tidal features that were later traced as a $\sim$5$^\circ$ stream spanning both sides of the core \citep{Shipp2018}. Spectroscopy confirmed a metal-poor stellar population ([Fe/H]$\,\approx -2.4$), an unresolved velocity dispersion, and a velocity gradient along the tails \citep{Simon2017, Li2018}. \citet{Shipp2018} also measured a distance gradient of $0.16 \pm 0.06$\,mag\,deg$^{-1}$ along the stream, with distance increasing toward the end nearer Sutlej on the sky, so that it spans $\sim$4\,kpc in distance, consistent with a radial orbit. Its \textit{Gaia} proper motion has been measured for both the core and the tails \citep{Shipp2019}, and modeling of its orbit indicates a recent close passage with the LMC, which has measurably perturbed the stream \citep{Erkal2018, Shipp2021}.

In this paper, we present the first deep photometric analysis of Sutlej in the third data release of the DECam Local Volume Exploration Survey (DELVE DR3) and investigate its association with Tuc III. Because DELVE reaches well below the main-sequence turnoff, where stream members vastly outnumber those on the giant branch accessible to \textit{Gaia}, it is sensitive to faint and extended structure that could not be detected in the discovery data. We use these data to recover the N24 branches and search for extensions beyond them, to measure the distance to each part of the stream independently from its main sequence and from standard candles, and to investigate whether Sutlej is associated with the other old, metal-poor substructure in the region, in particular Tuc III.

The paper is organized as follows. In Section~\ref{sec:data} we describe the DES, DELVE, and \textit{Gaia} datasets and the selection of our stellar sample. In Section~\ref{sec:results} we apply a matched filter, divide the stream into four segments, and measure each segment's distance from Hess diagrams and from standard candles. In Section~\ref{sec:tuc3} we test whether Sutlej could be tidal debris from Tuc III with restricted $N$-body simulations, and we summarize our conclusions in Section~\ref{sec:conclusions}. Appendix~\ref{app:sutlej_origins} discusses the river for which the stream is named.

\section{Data}
\label{sec:data}
Our photometry comes from DELVE DR3 (Section~\ref{sec:delve}), which incorporates the DES Y6 Gold catalog within the DES footprint. Our analysis region lies almost entirely within that footprint, with only 3.8 percent of stars coming from DELVE data outside DES after filtering our stellar sample (Section~\ref{sec:prep}). We use \textit{Gaia} DR3 astrometry (Section~\ref{sec:gaia}) to select standard candles.

\subsection{DES Y6 Gold and DELVE DR3}
\label{sec:delve}
\label{sec:desy6}
The DES Year 6 Gold catalog \citep{Bechtol2025} is a curated photometric dataset built from DES Data Release 2 \citep{Abbott2021}. It covers $\sim$5000\,deg$^2$ of the southern Galactic cap in the $grizY$ bands, imaged with the Dark Energy Camera \citep[DECam;][]{Flaugher2015} on the V\'ictor M.\ Blanco 4-m Telescope at the NSF Cerro Tololo Inter-American Observatory (CTIO), with median $10\sigma$ point-source depths of $g = 24.7$, $r = 24.4$, $i = 23.8$, $z = 23.1$, and $Y = 21.7$ \citep{Abbott2021}.

DELVE \citep{DrlicaWagner2021} provides contiguous coverage of the high-Galactic-latitude ($|b| > 10^\circ$) southern ($\delta < 30^\circ$) sky in the $griz$ bands, combining newly acquired DECam observations with all public archival DECam data. These data are processed with the DES Data Management pipeline \citep{Morganson2018}, with image detrending and coaddition closely following DES Y6, so that the DELVE and DES catalogs are consistent. Outside the DES footprint, DELVE DR3 covers an effective unmasked area of $\sim$12{,}000\,deg$^2$ to median $10\sigma$ point-source depths of $g = 24.2$, $r = 23.7$, and $i = 23.2$, with a larger scatter than DES ($\sim$0.5\,mag, compared to $\sim$0.2\,mag), in part because it combines many observing programs \citep{Tan2025census}. The DELVE DR3 Gold catalog (Drlica-Wagner et al., in preparation), built from these image coadds, incorporates the DES Y6 Gold catalog within the DES footprint, and the merged catalog covers $\sim$20{,}000\,deg$^2$ \citep{Tan2025sats}. Henceforth, ``DELVE DR3'' refers to this merged catalog.

\subsection{Gaia DR3}
\label{sec:gaia}
The third data release of the \textit{Gaia} mission \citep{GaiaDR3} contains $\sim$1.8 billion sources to a limiting magnitude of $G \approx 21$, of which $\sim$1.47 billion have parallaxes and proper motions from the Early Data Release 3 astrometric solution \citep{Lindegren2021}. At the distance of Sutlej, this limit corresponds to the red giant branch and horizontal branch, so \textit{Gaia} astrometry is available only for the brightest stream members. We use the DR3 proper motions and parallaxes to identify likely Sutlej members among the blue horizontal branch (BHB) and RR Lyrae (RRL) stars used as standard candles in Section~\ref{sec:candles} and to reject nearby foreground stars. The RRL stars are drawn from the catalog of \citet{Li2023}, which is a cleaned sample of $\sim$136{,}000 RRL stars identified in \textit{Gaia} DR3 \citep{Clementini2023}, with distances.

\subsection{Stellar Sample}
\label{sec:prep}
In this analysis, we use the extinction-corrected PSF magnitudes from DELVE DR3\footnote{Columns \texttt{PSF\_\allowbreak MAG\_\allowbreak APER\_\allowbreak 8\_\allowbreak G\_\allowbreak CORRECTED} and \texttt{PSF\_\allowbreak MAG\_\allowbreak APER\_\allowbreak 8\_\allowbreak R\_\allowbreak CORRECTED}.} and restrict the sample to stars with $16.0 < g < 24.0$ in magnitude, $-0.5 < g-r < 1.0$ in color, and \texttt{EXT\_XGB} $= 0$ for star--galaxy separation. Throughout this paper, $g$ and $r$ denote these extinction-corrected magnitudes. The faint limit follows \citet{Tavangar2022}, who adopted $g < 24$ for their DES Y6 stellar sample of the Phoenix stream, so that main-sequence stars at the limit still have well-measured $g - r$ colors. 

We work in the Magellanic Stream coordinate system $(L_{\rm MS}, B_{\rm MS})$ of \citet{Nidever2008}, as implemented in \texttt{gala} \citep{PriceWhelan2017}, in which the segments of Sutlej run at a shallow angle to lines of constant $B_{\rm MS}$. Our analysis region is $-40^\circ < L_{\rm MS} < -20^\circ$, $-18^\circ < B_{\rm MS} < -5^\circ$. There are 856{,}876 DELVE stars that pass our cuts in this region.

We also cross-match DELVE DR3 with \textit{Gaia} DR3 within a radius of $1.0''$, creating a catalog of objects that have a counterpart in both catalogs. For matched objects, we append the \textit{Gaia} proper motions, their uncertainties and correlation, and the parallaxes to the DELVE photometry. 441{,}518 DELVE DR3 stars that have a \textit{Gaia} DR3 counterpart with a proper motion are added to the cross-matched catalog, which we refer to as DELVE$\times$\textit{Gaia}.

\begin{figure}
  \centering
  \includegraphics[width=\columnwidth]{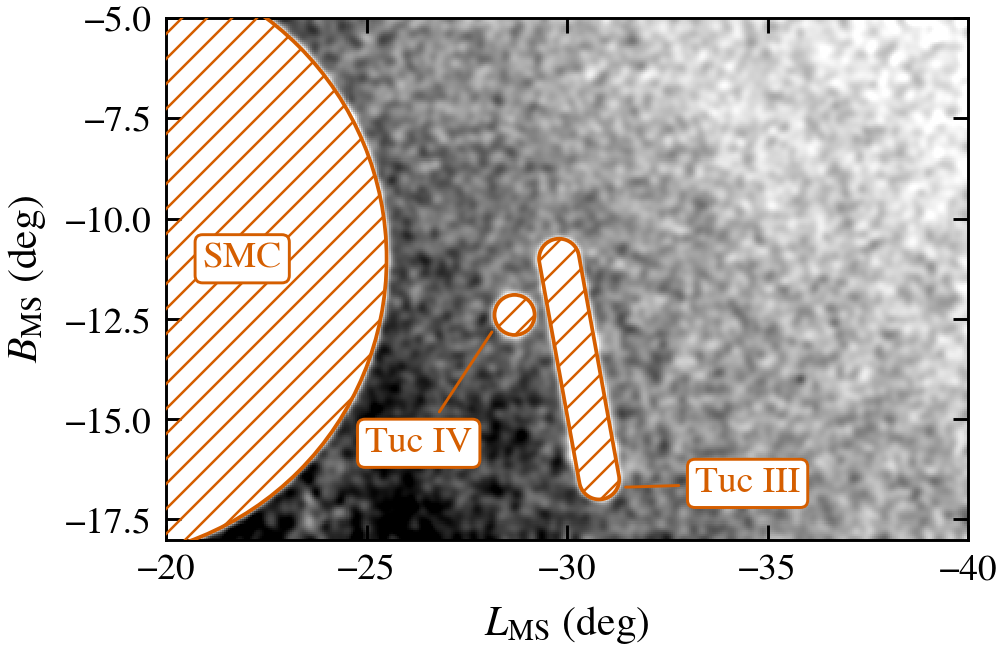}
  \caption{Smoothed density of all DELVE DR3 stars with $g<24$ in the analysis region ($0.05^\circ$ pixels, $0.1^\circ$ Gaussian smoothing). The hatched regions show the masked SMC, Tuc IV, and Tuc III stream regions.}
  \label{fig:region}
\end{figure}

To prevent the much denser stellar populations of systems in close proximity to Sutlej from dominating the matched-filter maps and background estimates, we mask three regions (Figure~\ref{fig:region}):
\begin{itemize}
  \item the SMC: a circle of radius $7.5^\circ$ centered on $(L_{\rm MS}, B_{\rm MS}) = (-18^\circ, -11^\circ)$;
  \item Tuc IV: a circle of radius $0.5^\circ$ centered on $(L_{\rm MS}, B_{\rm MS}) = (-28.69^\circ, -12.40^\circ)$;
  \item the Tuc III stream: all points within $0.5^\circ$ of the segment from $(L_{\rm MS}, B_{\rm MS}) = (-29.8^\circ, -11.0^\circ)$ to $(-30.8^\circ, -16.5^\circ)$.
\end{itemize}

Each mask radius is several times the extent of its corresponding system: the SMC's stellar density follows an exponential profile out to $\sim$7.5$^\circ$ \citep{Nidever2011}, the Tuc IV mask spans $\sim$3 times its 9.1$'$ half-light radius \citep{DrlicaWagner2015}, and the Tuc III mask extends $\sim$3 times the stream's 0.18$^\circ$ Gaussian width on either side of its track \citep{Shipp2018}.

After the masks and the $g < 24$ cut, 549{,}168 DELVE DR3 stars remain. Of these, 328{,}646 have a \textit{Gaia} DR3 counterpart, and 301{,}343 of those have \textit{Gaia} proper motions.

\section{Results}\label{sec:results}
We first build a main-sequence matched filter from the N24 isochrone and show that the same isochrone describes Tuc III (Section~\ref{sec:mf}). We then use the filtered maps to divide Sutlej into four linear segments for the purpose of analysis (Section~\ref{sec:segments}) and fit the distance of each segment, and of its two halves, from background-subtracted Hess diagrams (Section~\ref{sec:hess}). Finally, we check these distances with BHB and RRL stars selected by proper motion, compare that selection with the Tuc III stream, and use those stars to  estimate the stellar mass of Sutlej (Section~\ref{sec:candles}).

\subsection{Matched Filter}\label{sec:mf}

N24 found that Sutlej is well described by an old, metal-poor isochrone (10\,Gyr, $Z = 0.0003$, corresponding to [Fe/H]$\,\approx -1.8$) at $\sim$32\,kpc. We build a matched filter from a \citet{Marigo2017} isochrone with these parameters in the DES $g$ and $r$ bands. We also tested a MESA Isochrones and Stellar Tracks \citep[MIST;][]{Dotter2016, Choi2016} isochrone with the same age and metallicity and found consistent results. The filter selects stars whose $g-r$ color lies between $-0.05$ and $+0.07$\,mag of the isochrone color, widened at each magnitude by $2\,\sigma_g$, with $\sigma_g(g) = 0.001 + e^{(g-27.09)/1.09}$, which \citet{Shipp2018} fit to the median $g$-band uncertainty in DES. The uncertainty is evaluated $\Delta(m-M)/2 = 0.25$\,mag brighter for the blue edge and $0.25$\,mag fainter for the red edge \citep[$\Delta(m-M) = 0.5$\,mag;][]{Shipp2018}. The selection widens toward fainter magnitudes where photometric errors grow. We restrict the filter to the main sequence ($M_\mathrm{g} > 3.4$), since the many main-sequence stars below the turnoff dominate the sensitivity of DELVE to stream features, while the sparser giant branch is heavily contaminated by foreground stars.

\begin{figure}
  \centering
  \includegraphics[width=\columnwidth]{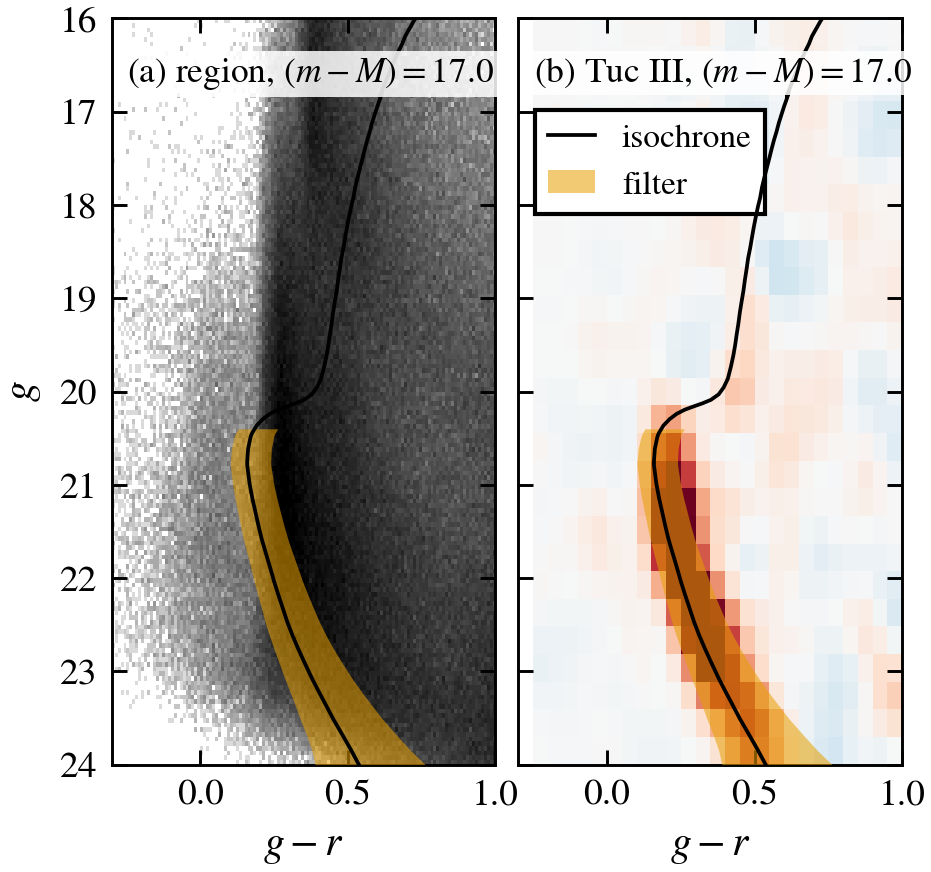}
\caption{(a) CMD of the masked analysis region, with the Sutlej isochrone (10 Gyr, Z=0.0003, \citealp{Marigo2017}) in black, and the main-sequence matched filter (orange) at $(m-M)=17.0$, adopted for Tuc III. (b) Background-subtracted Hess diagram of Tuc III, with the same isochrone and matched filter at the same distance modulus.}
  \label{fig:mf}
\end{figure}

We also compare the isochrone to Tuc III. Figure~\ref{fig:mf}a shows the filter overlaid on the CMD of the analysis region at $(m-M) = 17.0$ (25.1\,kpc), consistent with Tuc III's literature distance of 25 kpc \citep{DrlicaWagner2015}. Figure~\ref{fig:mf}b shows a background-subtracted Hess diagram of Tuc III with the Sutlej isochrone overlaid. With no adjustment to its age or metallicity, the isochrone follows the Tuc III main sequence and turnoff closely, so the two populations are consistent with the same age and metallicity. We return to the relation between Sutlej and Tuc III in Section~\ref{sec:tuc3}.

\subsection{Sutlej Segments}\label{sec:segments}
\begin{figure}
\centering
\includegraphics[width=\columnwidth]{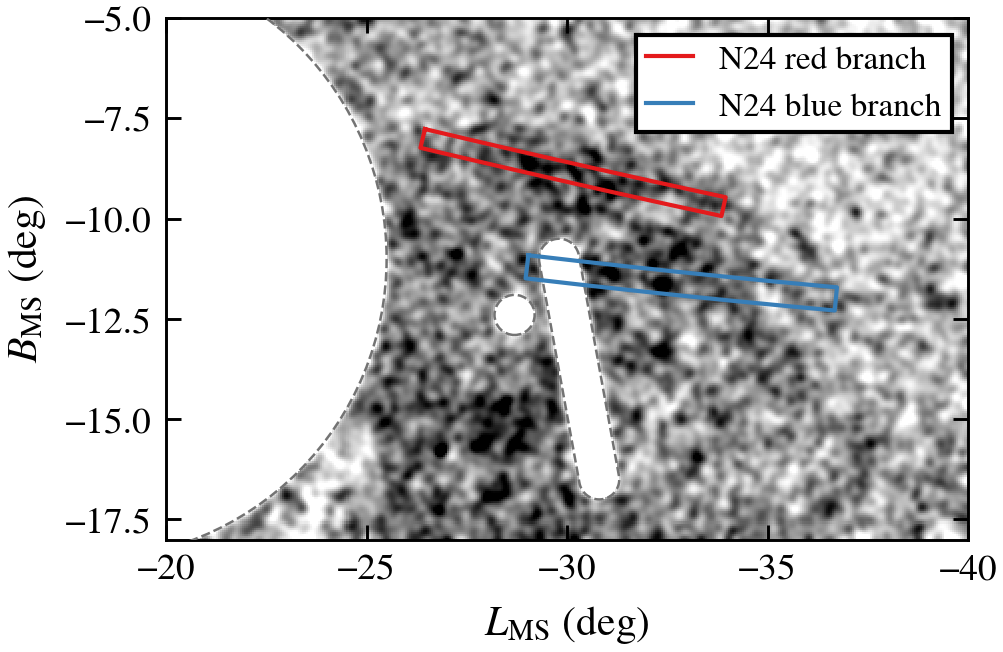}\caption{DELVE DR3 matched-filter map at $(m-M) = 17.5$, adopted from N24 for Sutlej (0.05$^\circ$ pixels, 0.1$^\circ$ Gaussian smoothing). Darker shading indicates higher filtered star counts. The red and blue boxes are the two branches of Sutlej identified by N24, drawn with half-widths of 0.24$^\circ$ and 0.29$^\circ$, respectively.}
\label{fig:n24}
\end{figure}

Figure~\ref{fig:n24} shows the DELVE DR3 matched-filter map at $(m-M) = 17.5$ with the two N24 branches overplotted. Both branches appear as clear overdensities. The map also shows an overdense feature running between the two branches, and structure that may continue from the bottom (blue) branch both toward and away from the SMC.

Because the stream is clearly not a single smooth track, we describe it as a set of linear segments, each with its own track. We place each segment by eye on the matched-filter maps at a range of distance moduli, since each segment is most distinct at a different distance. We define four segments, numbered 1--4, by their endpoints in $(L_{\rm MS}, B_{\rm MS})$:
\begin{enumerate}
  \item $(-26.0^\circ, -7.125^\circ) \to (-36.0^\circ, -10.625^\circ)$
  \item $(-30.5^\circ, -10.875^\circ) \to (-37.0^\circ, -13.875^\circ)$
  \item $(-27.0^\circ, -8.375^\circ) \to (-37.0^\circ, -12.375^\circ)$
  \item $(-26.0^\circ, -10.875^\circ) \to (-32.0^\circ, -11.625^\circ)$
\end{enumerate}

\begin{figure*}
  \centering
  \includegraphics[width=\textwidth]{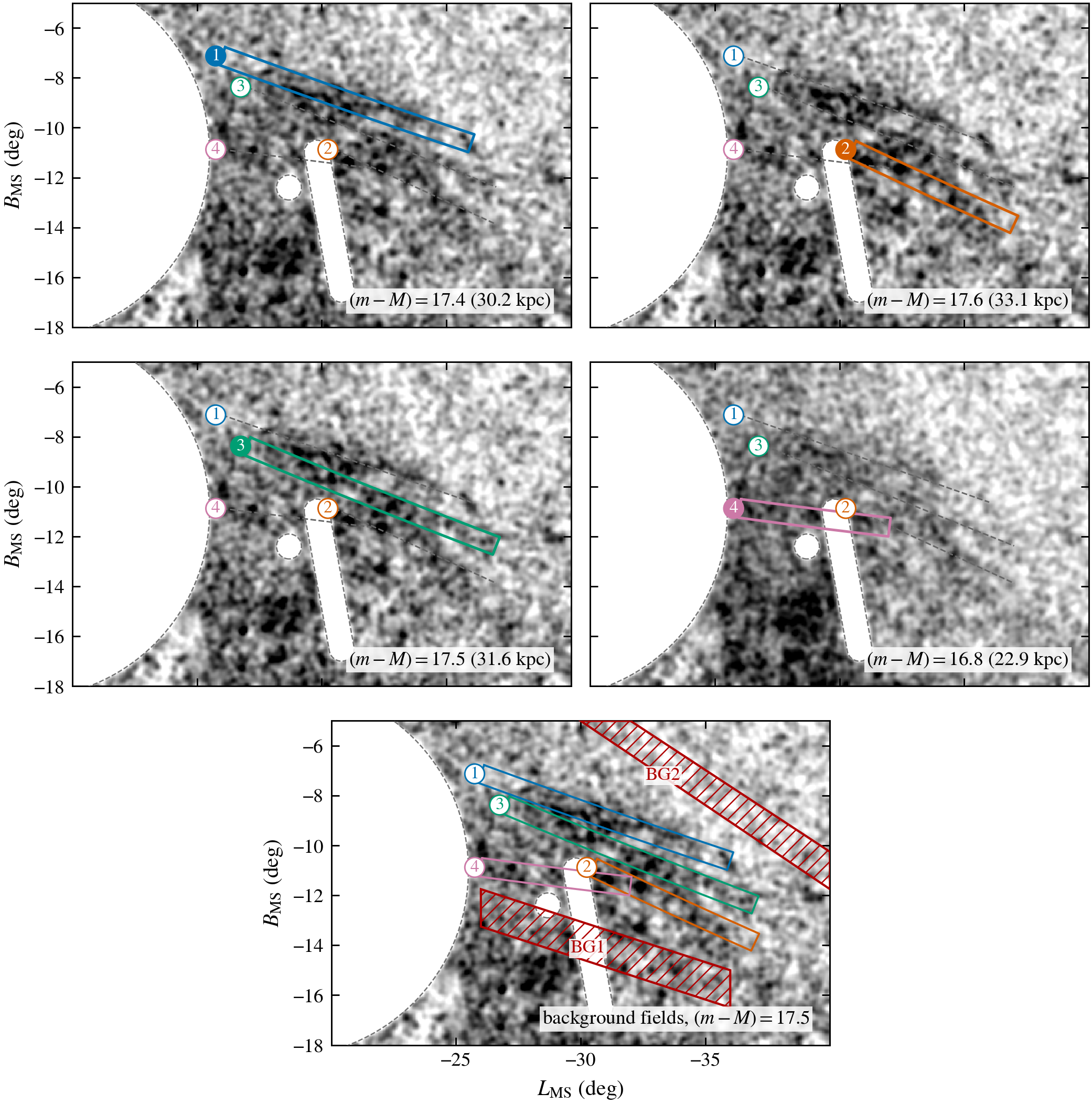}\caption{Matched-filter maps at the distance modulus of each segment (labeled with the corresponding distance) in $0.1^\circ$ bins smoothed with a Gaussian kernel of $\sigma = 0.15^\circ$. In each panel the segment fitted at that $(m-M)$ is outlined, and the other segments are shown as gray dashed lines. The bottom panel shows the map at $(m-M) = 17.5$ with all four segment boxes outlined and the two background fields (BG1 and BG2, red hatched) used for every background-subtracted Hess diagram. An animated version of this figure, stepping the matched filter through $16 \le (m-M) \le 19$ in 0.1\,mag steps with the segment boxes overlaid, is available at \href{https://doi.org/10.5281/zenodo.24207663}{doi:10.5281/zenodo.24207663} \citep{Burdett2026sweep}.}
  \label{fig:mfmu}
\end{figure*}

For each segment, we define a great-circle frame $(\phi_1, \phi_2)$ whose equator passes through the two endpoints, and take the segment's box to be $|\phi_2| \le 0.375^\circ$ between the endpoints (Figure~\ref{fig:mfmu}). Segments 1 and 3 straddle the N24 red branch, with segment 3 corresponding to the overdensity between the two N24 branches. Segment 2 follows the N24 blue branch and extends it $\sim$3$^\circ$ lower in $L_{\mathrm{MS}}$ and in $B_{\mathrm{MS}}$. Segment 4 extends the blue branch toward the SMC along its extrapolated track, ending at $(L_{\rm MS}, B_{\rm MS}) = (-26^\circ, -10.9^\circ)$.

\subsection{Segment Distances}\label{sec:hess}
\begin{figure*}
  \centering
  \includegraphics[width=\textwidth]{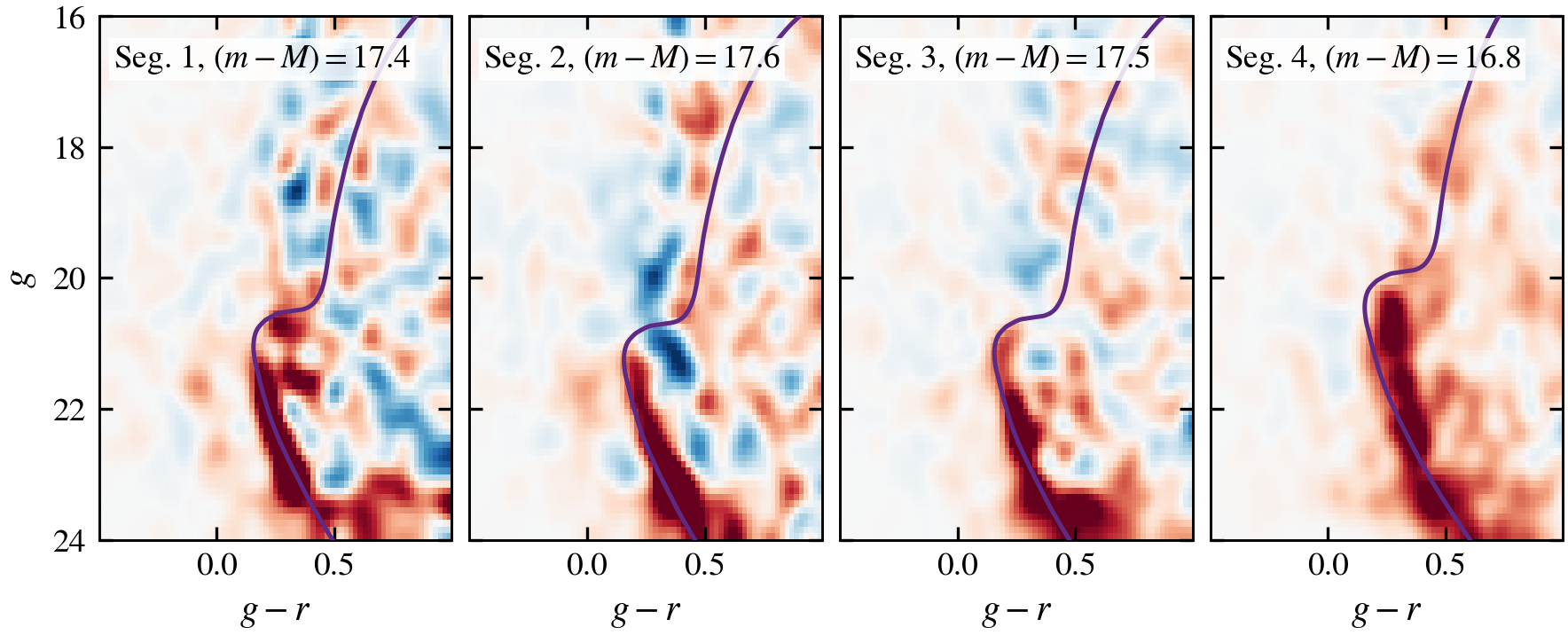}\caption{Hess diagrams of the four segments (red: excess over background). Purple: the Sutlej isochrone at each segment's by-eye distance modulus $(m-M)$.}
  \label{fig:hess}
\end{figure*}

For each segment, we build a background-subtracted Hess diagram, taking the stars within the segment box as the on-stream sample. The background comes from two fixed off-stream fields parallel to the segments: BG1, a strip below segment 2 with vertices $(L_{\rm MS}, B_{\rm MS}) = (-26.0^\circ, -13.25^\circ)$, $(-36.0^\circ, -16.5^\circ)$, $(-36.0^\circ, -15.0^\circ)$, and $(-26.0^\circ, -11.75^\circ)$, and BG2, a strip at higher $B_{\mathrm{MS}}$ than segment 1 with vertices $(-32.0^\circ, -5.0^\circ)$, $(-40.0^\circ, -10.25^\circ)$, $(-40.0^\circ, -11.75^\circ)$, and $(-30.0^\circ, -5.0^\circ)$. These fields sample the surroundings of the stream while avoiding the masked regions and other known substructure, and the background is scaled to the on-stream sample by the ratio of unmasked areas, measured on a HEALPix ($N_{\rm side}=512$) grid. The Hess diagrams use 75 color bins over $-0.5 < g-r < 1.0$ and 80 magnitude bins over $16 < g < 24$, smoothed with a Gaussian of width two bins. All four segments show a clear main sequence and turnoff traced by the Sutlej isochrone (Figure~\ref{fig:hess}). We fit the distance modulus of each segment by eye, shifting the isochrone in steps of 0.1\,mag. This gives $(m-M) = 17.4$, 17.6, 17.5, and 16.8 for segments 1--4 (30.2, 33.1, 31.6, and 22.9\,kpc), and each segment is most distinct in the matched-filter map at its fitted distance modulus (Figure~\ref{fig:mfmu}). These by-eye fits should be treated as preliminary estimates with significant uncertainty. With constant age and metallicity, the absolute distances carry a systematic uncertainty of $\sim$0.1\,mag from the isochrone model and assumed stellar population \citep{Cerny2023}. The segment distances would also be essentially unchanged with an isochrone matching the literature parameters of Tuc III (13\,Gyr, $Z = 0.0001$; \citealt{Shipp2018}), whose metallicity is close to the spectroscopic value of $\mathrm{[Fe/H]} \approx -2.4$ \citep{Simon2017, Li2018}. Below the turnoff, the two isochrones differ by less than 0.02\,mag in $g-r$, and the offset that best aligns their main sequences is 0.01\,mag in $(m-M)$. To look for distance gradients along each segment, we repeat the fit separately for the left and right halves of each segment, split at its midpoint; the left/right values are 17.3/17.5, 17.7/18.2, 17.5/17.8, and 16.8/17.5 for segments 1--4, respectively (Appendix~\ref{app:halves}).

\subsection{Standard Candles}\label{sec:candles}
\begin{figure}
  \centering
  \includegraphics[width=\columnwidth]{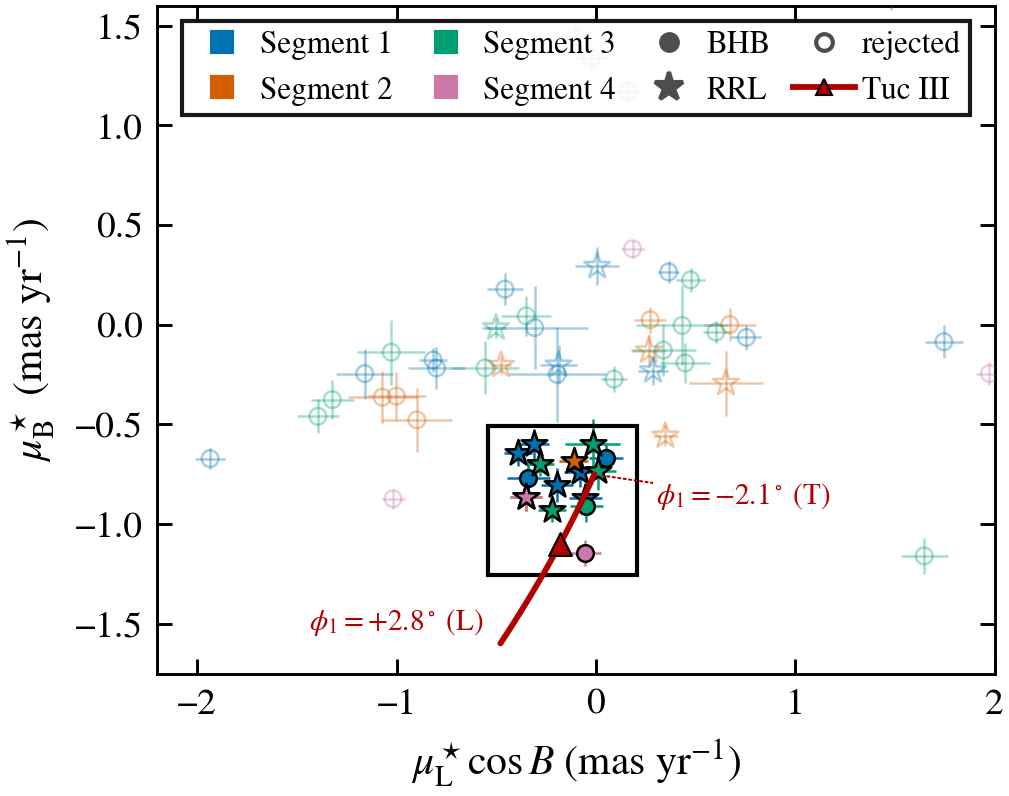}
  \caption{Reflex-corrected proper motions of candidate BHB (circles) and RRL (stars) members in the four segment boxes, with $1\sigma$ error bars. Filled symbols are stars whose proper motions lie within the selection box (black rectangle) to within their $1\sigma$ uncertainties; open symbols are rejected. One RRL lies in both the segment 2 and 4 boxes and is shown in both. 64 rejected entries lie outside the plotted window. The red line is the reflex-corrected proper motion of the Tuc III stream between the end points of \citet{Shipp2018} ($-2.1^\circ \le \phi_1 \le +2.8^\circ$), from the proper motions of \citet{Shipp2019}, a distance of 25\,kpc at the core \citep{DrlicaWagner2015}, and $d(m-M)/d\phi_1 = -0.16$\,mag\,deg$^{-1}$ \citep{Shipp2018}. The triangle marks the core; T and L label the trailing ($\phi_1 < 0$) and leading ($\phi_1 > 0$) ends.}
  \label{fig:pm}
\end{figure}

We also identify BHB and RRL stars in each segment box.
\begin{itemize}
  \item \textbf{BHBs} are DELVE$\times$\textit{Gaia} stars on the empirical M92 BHB locus \citep{Belokurov2007}, transformed to DES photometry \citep{Li2019} and shifted from $(m-M)_{\rm M92} = 14.59$ \citep{Harris1996} to the segment's distance modulus. Their individual distances come from the $M_\mathrm{g}$--color relation of \citet{Barbosa2022}.
  \item \textbf{RRLs} are taken from \citet{Li2023}. We keep those with $|(m-M) - (m-M)_{\mathrm{segment}}| \le 1$\,mag.
\end{itemize}
To isolate stream members among these candidates, we also require proper motions consistent with Sutlej. We correct each star for solar reflex motion at its own standard-candle distance and rotate the result into $(\mu^\star_{\rm L}\cos B, \mu^\star_{\rm B})$, where $\star$ denotes reflex-corrected proper motions. We derive the selection box from the proper-motion range of the N24 members ($0.9 < \mu_{\rm L}\cos B < 1.9$, $-0.8 < \mu_{\rm B} < 0.2$\,mas\,yr$^{-1}$ as observed), reflex-correcting its center at $(m-M) = 17.5$ at the center of the segment region and tightening it to $-0.54 < \mu^\star_{\rm L}\cos B < 0.21$, $-1.26 < \mu^\star_{\rm B} < -0.51$\,mas\,yr$^{-1}$; Figure~\ref{fig:pm}). A star is accepted if its proper motion lies within the box to within its $1\sigma$ uncertainties in each component. Of 121 candidate BHB and RRL stars with \textit{Gaia} proper motions inside the distance windows, 15 pass (7, 2, 5, and 2 in segments 1--4 with 1 overlap. The accepted stars cluster tightly in reflex-corrected proper motion, and the rejected ones scatter broadly.

\begin{figure*}
  \centering
  \includegraphics[width=\textwidth]{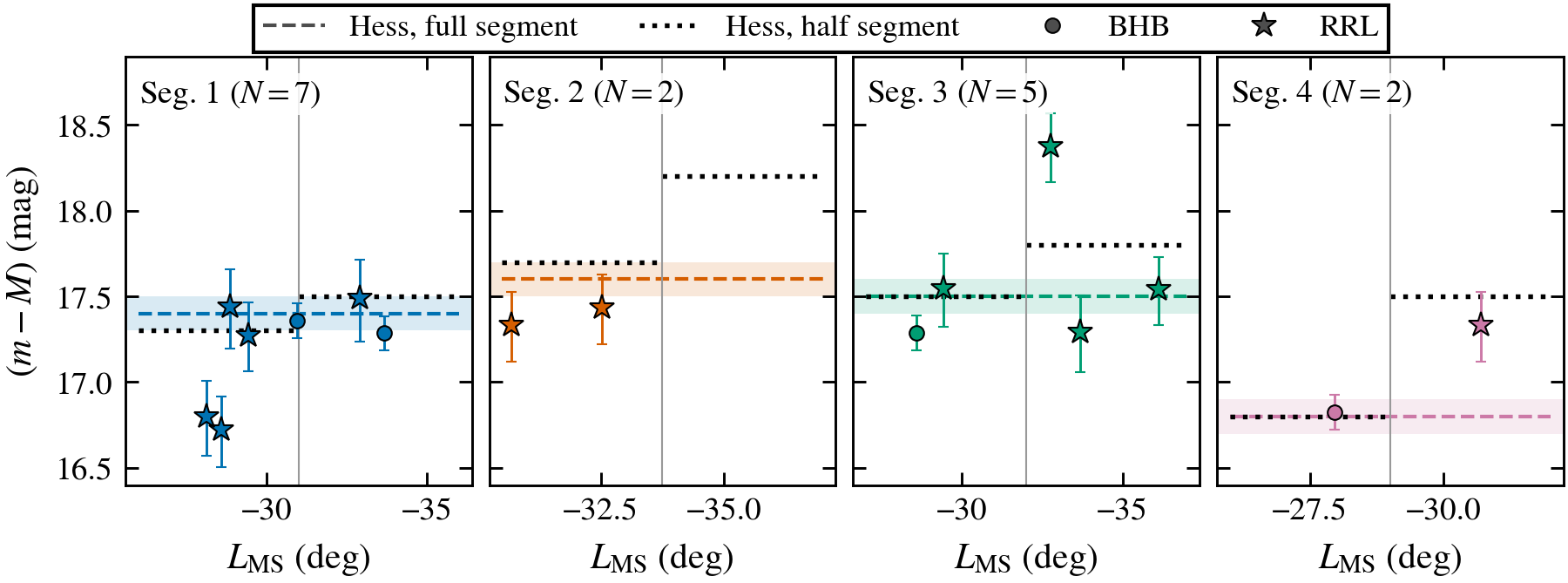}
  \caption{Distance moduli of the accepted BHB (circles) and RRL (stars) members along each segment, with error bars for the uncertainty in distance. The dashed line is the full-segment Hess fit $(m-M)$, with a 0.1 magnitude band (shaded), and the dotted black lines are the half-segment Hess fits. The gray vertical line marks the midpoint of each segment, and $N$ is the number of accepted stars.}
  \label{fig:dist}
\end{figure*}

Figure~\ref{fig:dist} compares the distances of the accepted BHB and RRL stars with the Hess-diagram fits. The BHBs are shown with a 0.1 magnitude uncertainty in distance modulus (following \citealt{Deason2011}), and the RRLs are shown with the uncertainties given in the \citet{Li2023} catalog. Most agree with the full-segment $(m-M)$ within this uncertainty. Given the small number of members per segment, we cannot fit a distance gradient from the standard candles alone, but their distances are generally consistent with the direction of the gradients inferred from the left and right half-segment Hess diagrams. The half-segment values hint at an overall increasing trend in distance toward lower $L_{\mathrm{MS}}$ (right), although three of four segments are consistent with a flat per-segment distance gradient (Figure~\ref{fig:dist}).

Because Sutlej passes within a few degrees of Tuc III, we compare our selection with the Tuc III stream (Figure~\ref{fig:pm}, red). We take the proper motion of the stream and its gradient along $\phi_1$ from \citet{Shipp2019}, in their Tuc III frame, in which the $\phi_1 < 0$ arm lies at higher $B_{\rm MS}$, on the side nearer Sutlej. We take its extent, $-2.1^\circ \le \phi_1 \le +2.8^\circ$, and distance gradient, $d(m-M)/d\phi_1 = -0.16$\,mag\,deg$^{-1}$, from \citet{Shipp2018}, about a core distance of $(m-M) = 17$. The distance gradient means the $\phi_1 < 0$ (trailing) arm is also the more distant one. At its trailing end ($L \approx -29.8^\circ$, $B \approx -10.8^\circ$) Tuc III reaches $(m-M) \approx 17.3$ (29\,kpc). That end lies within $\sim$0.5$^\circ$ of the right half of segment 4, at $(m-M) = 17.5$ (31.6\,kpc), and $\sim$0.7$^\circ$ from the start of segment 2, whose left half is at $(m-M) = 17.7$ (34.7\,kpc). Where the two streams are closest on the sky, they are therefore also within $\sim$0.2--0.4\,mag in distance, much closer than the 0.5--0.6\,mag between the Tuc III core and Sutlej. Their reflex-corrected proper motions also overlap there. For $-2.1^\circ \le \phi_1 \lesssim +0.9^\circ$, i.e.\ the trailing arm and the core, the Tuc III track lies inside the Sutlej selection box, and its trailing end falls among the accepted stars. Neither proper motion nor the $\pm1$\,mag distance windows can therefore separate Sutlej members from Tuc III debris near the trailing arm.

The accepted stars also give a rough stellar mass for Sutlej. Following \citet{Ferguson2022}, we convert the 15 unique HB stars to a stellar mass using the number of HB stars expected per unit mass for an old, metal-poor population with a \citet{Chabrier2001} IMF, $\sim$17 per $10^{4}\,\mathrm{M}_\odot$. This gives $M_\star = 9.4^{+2.6}_{-2.2} \times 10^{3}\,\mathrm{M}_\odot$ in living stars, about 3.5 times lower than the $3.3 \times 10^{4}\,\mathrm{M}_\odot$ estimated by N24.

\section{A Tuc III Origin for Sutlej?}\label{sec:tuc3}

We now consider whether Sutlej could be associated with Tuc III. The two systems lie within a few degrees of each other at similar distances, and the reflex-corrected proper motions of Tuc III members in the area of Sutlej \citep{Shipp2019} are consistent with the Sutlej BHB and RRL stars found in Section~\ref{sec:candles}. However, the Tuc III tails run nearly perpendicular to the Sutlej segments, and unperturbed debris following the Tuc III orbit would not be expected to reach the Sutlej field.

The recent orbital history of Tuc III offers a possible explanation. Tuc III had a close encounter with the LMC $\lesssim 100$\,Myr ago \citep{Shipp2021}, which substantially deflected its orbit \citep{Erkal2018}. Accounting for this encounter when integrating the orbit backward places the preceding pericenter with the Milky Way, $\sim$0.5\,Gyr ago, at $\sim$1.5\,kpc or less. At such a small pericenter, the inner Galaxy delivers a strong tidal shock that can unbind a large fraction of the progenitor's stars in a single passage \citep{Gnedin1999}, and the subsequent LMC encounter can then perturb this extended debris away from the Tuc III orbital track. We use simulations to test whether these processes could produce debris resembling Sutlej.

\subsection{Simulations}\label{sec:tuc3sims}

Because a strong pericentric tidal shock strips most of the progenitor over a short interval, we model the disruption with restricted $N$-body simulations, implemented in \texttt{nbody\_streams}\footnote{\url{https://github.com/appy2806/Nbody_streams}} \citep{Arora2026}, rather than with a particle-spray model, which assumes gradual stripping. The Tuc III progenitor consists of an NFW halo with a subdominant \citet{Plummer1911} stellar component, as in other models of tidally disrupting dwarf galaxies \citep[e.g.,][]{Limberg2026, Li2026}. Each simulation integrates the present-day phase-space position of Tuc III backward to the start time, centers the progenitor's particles on the orbit at that time, and evolves them forward with the satellite's self-gravity computed from its particles. The host potential combines the \citet{McMillan2017} Milky Way, as realized in \citet{Shipp2021}, with the non-axisymmetric part of the rotating \citet{Sormani2022} bar and the NFW LMC of \citet{Vasiliev2021}, whose orbit is integrated under the mutual gravity of the two galaxies.  In this model the LMC and Milky Way dark matter halos remain rigid.

We model the progenitor as a dark-matter-dominated dwarf galaxy, a choice set by the limits of a collisionless scheme rather than a conclusion about the nature of Tuc III, which remains debated. \citet{Simon2017} favored a dwarf galaxy on the basis of its large size and low metallicity relative to globular clusters of similar luminosity, and \citet{Li2018} found a possible metallicity spread among stars in the tails. However, its large size may itself be a consequence of tidal disruption, and its velocity dispersion is only an upper limit. The $r$-process enhancement of its brightest members, rarely seen in ultra-faint dwarfs \citep{Hansen:2017, Marshall2019}, may instead point to a globular cluster, and \citet{Zaremba:2025} have argued from a re-examination of the evidence that Tuc III may be a star cluster \citep[see also Appendix D of][]{Cerny:2026}. A cluster progenitor would require modeling collisional dynamics, but we expect the large-scale distribution of the debris to be set mainly by the orbit and the LMC encounter rather than by the internal structure of the progenitor.

We generate 500 realizations that sample the observational uncertainties on Tuc III, the bar and LMC parameters, and the properties of the progenitor (Appendix~\ref{app:tuc3}). The progenitor's structure at the start of each simulation is left free and constrained only by requiring a bound remnant that matches Tuc III today in size and velocity dispersion, so the halo at the start of the simulations may represent a dwarf galaxy that has already been stripped by previous pericentric passages. The halo mass within 37\,pc, the half-light radius of Tuc III \citep{MutluPakdil2018} at 25\,kpc, is drawn log-uniformly between $10^{4}$ and $3 \times 10^{5}\,\mathrm{M}_\odot$ and its scale radius between 50 and 500\,pc, corresponding to total halo masses of $\sim$10$^{5}$--10$^{8}\,\mathrm{M}_\odot$ at the start of the simulation. The stellar component is a Plummer sphere with a radius of 20--50\,pc and a nominal mass of $10^{3}\,\mathrm{M}_\odot$, light enough to trace the stripping without affecting it. Each model is sampled with 15{,}000 dark matter and 4{,}000 stellar particles and evolved forward from a start time between 0.8 and 3\,Gyr ago, which includes the most recent pericentric passage. These models test whether debris from a Tuc III-like progenitor can plausibly reach the Sutlej field; they are not fits to either system. Because the sampling is not designed to be representative, the fraction of models that resemble Sutlej should not be read as the probability of a Tuc III origin.

\subsection{Comparison with Sutlej}\label{sec:tuc3comp}

\begin{figure*}
\centering
\includegraphics[width=0.9\textwidth]{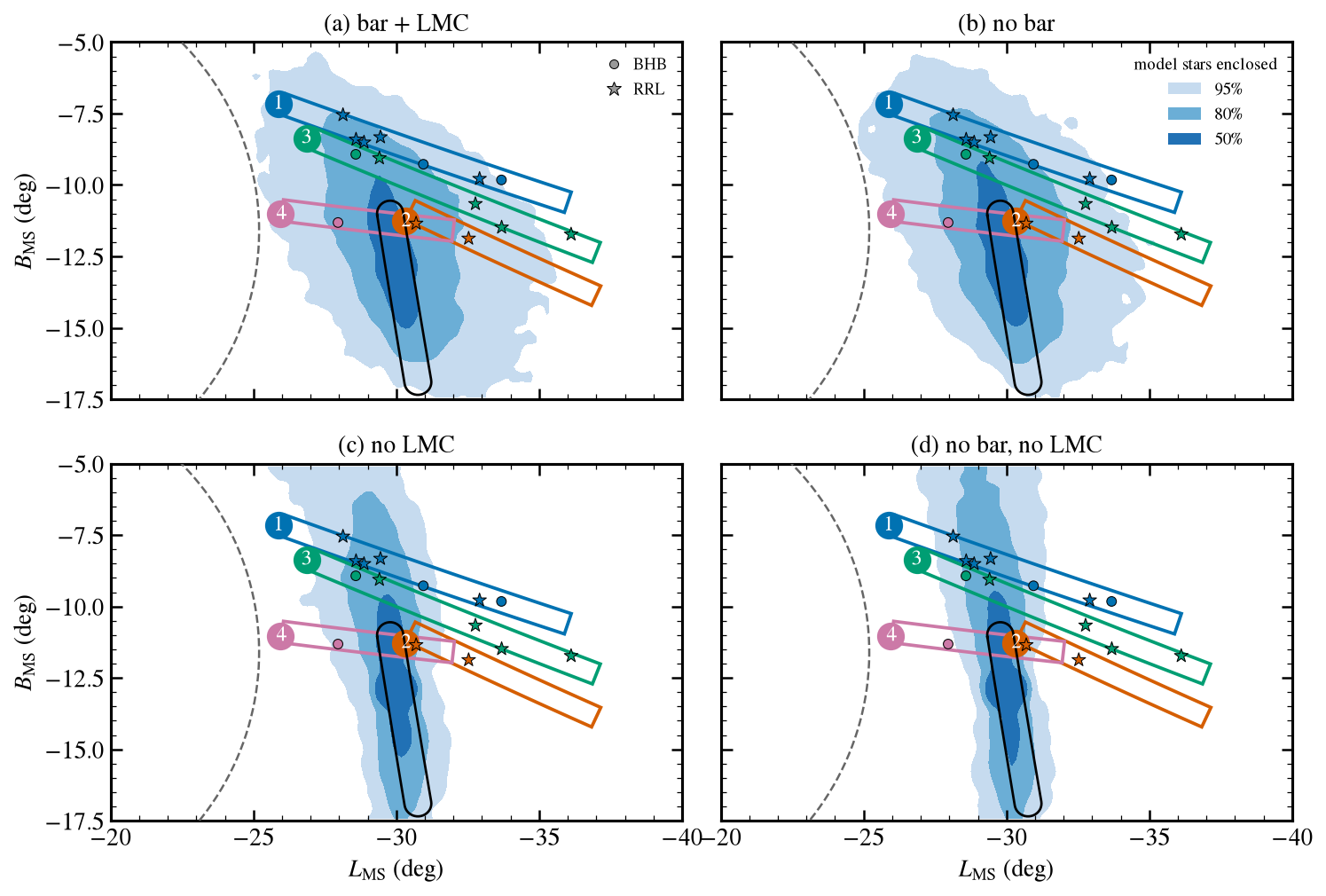}
\caption{Stacked distribution of model stellar debris with $16.7 < (m-M) < 17.8$ for the 24 Tuc III-consistent restricted $N$-body realizations in which every Sutlej segment is overdense away from the Tuc III track in the full potential, each weighted equally. Panels show the same models in (a) the full potential with bar and LMC, (b) no bar, (c) no LMC, and (d) neither. Filled contours enclose 95, 80, and 50 percent of the model stars in each panel (key in panel b). Colored boxes mark Sutlej segments 1--4, circles and stars mark the accepted BHB and RRL members respectively, the black outline marks the Tuc III stream, and the dashed arc marks the SMC mask. Without the LMC, the debris forms a narrow band along the Tuc III orbit; with it the envelope broadens across the track and covers the segments.}
\label{fig:tuc3model}
\end{figure*}

To isolate the effects of the bar and the LMC, we rerun each model in four host potentials: with both the bar and LMC, without the bar, without the LMC, and without either, recomputing the past orbit in each case. We then identify the models in which the debris resembles Sutlej by requiring, in the full potential, an overdensity in each of the four segment boxes relative to the surrounding field, evaluated away from the Tuc III track and at each segment's distance modulus (Appendix~\ref{app:sutlejsel}). Figure~\ref{fig:tuc3model} shows the stacked stellar debris of the selected models in each host potential, and three individual realizations are shown in Appendix~\ref{app:tuc3} (Figure~\ref{fig:tuc3examples}).

The simulations illustrate how the combination of a deep pericentric passage and the subsequent LMC encounter can spread Tuc III debris across the Sutlej field. The strong tidal field of the inner Galaxy strips a large fraction of the progenitor's stars at pericenter, producing debris with a broad range of orbital energies that extends several kpc along the orbit by the time of the LMC encounter. The LMC then differentially perturbs this extended debris, spreading it across the original orbital track. The bar has comparatively little effect on the resulting distribution. Without the LMC, the debris remains concentrated in a narrow band along the Tuc III orbit and overlaps the Sutlej segments primarily where they cross this band near $L_{\rm MS} \approx -30^\circ$. Including the LMC deepens the preceding pericenter, producing more extensive stripping and a broader debris distribution that reaches across the Sutlej field. The models whose debris reaches the Sutlej field are those with the deepest pericenters, typically $\lesssim$0.7\,kpc, and their present-day distances and proper motions lie close to the measured values for Tuc III (Appendix~\ref{app:tuc3}).

The simulated debris is also broadly consistent with the distances and proper motions of Sutlej (Figure~\ref{fig:tuc3examples}). The debris becomes more distant toward lower $L_{\rm MS}$, with the observed segment distances generally lying along the model distribution. This trend agrees with the half-segment fits for each segment. Within the segment boxes, the simulated debris occupies a narrow, elongated locus in proper motion that includes most of the candidate BHB and RRL members, with a spread of $\sim 0.1\,\mathrm{mas\,yr^{-1}}$, comparable to the \textit{Gaia} uncertainties. This kinematic coherence is compatible with the broad spatial distribution: a velocity difference of $\sim 10\,\mathrm{km\,s^{-1}}$ is small compared to the orbital velocity of $\sim 200\,\mathrm{km\,s^{-1}}$, but can produce separations of several kpc over $\sim 0.6\,\mathrm{Gyr}$, corresponding to angular scales of order $10^\circ$ at the distance of Sutlej.

Despite these similarities, the models do not reproduce the four distinct segments of Sutlej. Some realizations show enhanced density along the segments, but the debris generally forms a broad distribution rather than the narrow, separated structures seen in the DELVE data. The stellar mass budget is a further consideration. N24 estimated a stellar mass of $\sim 3 \times 10^4\,\mathrm{M}_\odot$ for Sutlej, roughly ten times the combined mass of the Tuc III core and known tails \citep[$\sim 3.4 \times 10^3\,\mathrm{M}_\odot$;][]{DrlicaWagner2015, Shipp2018}. Our rough estimate from HB stars (Section~\ref{sec:candles}) is $9.4\times10^{3}\,\mathrm{M}_\odot$, still greater than that of Tuc III. A shared progenitor would therefore have lost the large majority of its stars. Consistent with this picture, the models that place debris in the Sutlej field are among the most heavily stripped, retaining a median of $\sim$6 percent of their stellar particles in a bound remnant compared to $\sim$13 percent for all Tuc III-consistent models. Extensive mass loss is independently suggested by the spectroscopic metallicity of Tuc III \citep{Simon2017}, which places it above the stellar mass--metallicity relation of Milky Way satellites, more metal-rich than expected for its present stellar mass; in Auriga simulations, satellites in this regime have lost substantial stellar mass \citep{Riley2026}. \citet{Dominguez2026} find six RRL stars in the Tuc III tails and none in the remnant, which implies a lost stellar mass of $(2$--$34) \times 10^{4}\,\mathrm{M}_\odot$, a lower limit since older debris may have phase-mixed beyond the known tails. These results are consistent with a progenitor considerably more massive than the present-day Tuc III remnant, potentially comparable to the stellar mass estimated for Sutlej. However, the RRL-based mass loss, the inferred initial mass, and both stellar mass estimates of Sutlej are uncertain.

\begin{figure}
\centering
\includegraphics[width=\columnwidth]{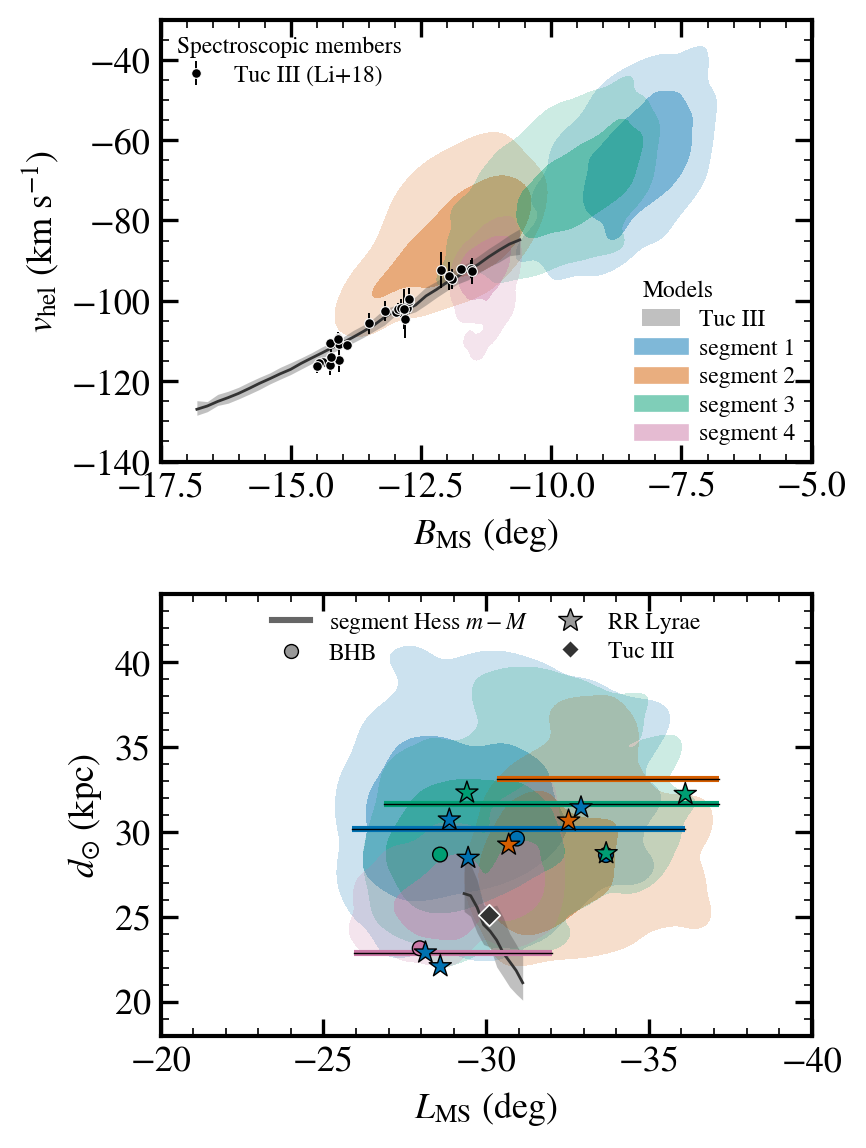}
\caption{Predicted kinematics and distances of Tuc III debris in the Sutlej segments for the 24 models in Figure~\ref{fig:tuc3model}. In both panels, colored contours (68 and 95 percent) show unbound model debris in each segment box, and the gray band is the 16th--84th percentile range of the model Tuc III stream.
\emph{Top:} Heliocentric radial velocity versus $B_{\rm MS}$, counting stars within $\pm 0.3$\,mag of each segment's distance modulus. Black points are the Tuc III members of \citet{Li2018}.
\emph{Bottom:} Heliocentric distance versus $L_{\rm MS}$, for all model debris in each segment box with no distance-modulus cut. Horizontal bars mark each segment's isochrone-fit distance and span its $L_{\rm MS}$ extent. Circles and stars are the BHB and RR Lyrae candidates, colored by segment, and the diamond marks Tuc III ($m-M = 17.0$).
The contours combine the spread within and between models and are intended only as an approximate guide.}
\label{fig:tuc3vr}
\end{figure}

The same models also predict the radial velocities of Sutlej, which have not yet been measured. Figure~\ref{fig:tuc3vr} shows the heliocentric radial velocities of model debris in each segment box as a function of $B_{\rm MS}$, together with the Tuc III stream in the same models and the spectroscopic members of \citet{Li2018}. The models reproduce the observed velocity gradient along the Tuc III tails, and the debris in the Sutlej segments continues this trend toward less negative velocities with increasing $B_{\rm MS}$. Segment 4, the segment nearest Tuc III on the sky and in distance, is predicted to lie close to the extension of the Tuc III velocity gradient, at $v_{\rm hel} \sim -100$ to $-85$\,km\,s$^{-1}$, while segment 2 spans $\sim -100$ to $-70$\,km\,s$^{-1}$ and segments 1 and 3, which lie farthest from the Tuc III track, are offset to $\sim -90$ to $-50$\,km\,s$^{-1}$. These ranges combine the spread within individual models with the variation between them and should be taken as approximate, but the predicted offsets of tens of km\,s$^{-1}$ between the segments, and between segments 1 and 3 and the Tuc III tails, are large compared to typical spectroscopic uncertainties and provide a direct test of the association.

Detailed debris morphology in the simulations may be affected by a number of factors. For example, we consider only a dark-matter-dominated progenitor, rather than exploring possible star-cluster progenitors. The Milky Way and LMC are represented by rigid potentials, and we adopt a single realization of the Milky Way mass distribution. In particular, the strength of the pericentric tidal shock depends on the inner Galactic potential, while the time evolution of the Galaxy can itself imprint structure on streams \citep{Arora2026}. These effects are not fully captured by the present models.

The simulations therefore demonstrate that debris from a Tuc III-like progenitor can reach the Sutlej field with distances and proper motions broadly consistent with the observations. Strong stripping at pericenter followed by differential perturbations from the LMC provides a possible explanation for the otherwise unexpected orientation and extent of the debris, although the two distinct N24 branches remain unexplained. Spectroscopic measurements of Sutlej members would provide a direct test of a shared origin, since a common progenitor requires radial velocities that follow the trend predicted in Figure~\ref{fig:tuc3vr} and metallicities consistent with those of Tuc III \citep[{[Fe/H]}\,$\approx -2.49$;][]{Li2018}. Additionally, if the LMC encounter scattered debris from both arms of the progenitor, comparable structure may lie at lower $B_{\rm MS}$ than the extent of our analysis region, and a search there would test whether Sutlej is one part of a broader Tuc III debris field.

\section{Conclusions}\label{sec:conclusions}
We present the first analysis of the Sutlej stellar stream in deep DECam photometry from DELVE DR3. We find that:
\begin{enumerate}
  \item We detect Sutlej in its main-sequence stars for the first time in DELVE DR3. N24 identified Sutlej in \textit{Gaia} DR3, which at $\sim$32\,kpc reaches only the red giant branch and horizontal branch, using a 10\,Gyr, $Z = 0.0003$ isochrone. Our sample extends to $g = 24$, $\sim$2.5--3.5\,mag below the main-sequence turnoff, and a main-sequence matched filter built from the same isochrone recovers both N24 branches.

  \item Sutlej appears to be more extended and with more structure than the two branches reported in N24. We describe it as four linear segments spanning $\sim$11$^\circ$ in $L_{\rm MS}$ and $\sim$7$^\circ$ in $B_{\rm MS}$. These segments include three features that \textit{Gaia} could not reveal: an overdensity between the two branches reported in N24 (segment 3), a continuation of the lower branch left toward the SMC (segment 4), and an extension of that branch $\sim$3$^\circ$ to the right (segment 2).

  \item We measure that the segments lie at different distances. Fits to their background-subtracted Hess diagrams give $(m-M) = 17.4$, 17.6, 17.5, and 16.8 (30.2, 33.1, 31.6, and 22.9\,kpc) for segments 1--4. Fits to the two halves of each segment suggest gradients of up to $\sim$0.7\,mag, with all segments more distant toward lower $L_{\mathrm{MS}}$. Fifteen BHB and RRL stars with Sutlej-like reflex-corrected proper motions provide a check on these distances, most agreeing with their segment's Hess fit to within 10 percent.

  \item Sutlej closely resembles Tuc III in on-sky position, color-magnitude space, and proper motion. The Sutlej isochrone fits the Tuc III main sequence and turnoff at its distance of 25\,kpc. The reflex-corrected proper motion of Tuc III falls within the Sutlej selection box. Near the trailing arm of Tuc III, the two streams lie within $\sim$0.5--0.7$^\circ$ on the sky and $\sim$0.2--0.4\,mag in distance, so Sutlej members are indistinguishable from Tuc III debris there.

  \item Restricted $N$-body simulations show that debris from a Tuc III-like progenitor can reach the Sutlej field. The LMC encounter $\lesssim$100\,Myr ago means that the preceding pericenter of Tuc III, $\sim$0.5\,Gyr ago, reached $\sim$0.3--1.5\,kpc, compared to $\sim$4\,kpc for the same orbits without the LMC. This passage strips most of the progenitor, and the LMC then spreads the debris across the original orbital track. Of 182 Tuc III-consistent models, 24 show overdensities in all four segments in the full potential, compared to 4 without the LMC. These models also follow the distance trend of Sutlej and occupy a narrow proper-motion locus that contains most of the BHB and RRL members. The bar has little effect compared to the LMC.

  \item The models do not reproduce distinct segments of Sutlej, and instead produce a broad distribution of debris. A shared origin would also require the progenitor to have lost the large majority of its stars, since Sutlej holds roughly 3 to 10 times the stellar mass of the Tuc III core and tails. Such heavy stripping is consistent with the position of Tuc III above the stellar mass--metallicity relation, but neither object's stellar mass is well constrained.

  \item The models make testable radial-velocity predictions. They reproduce the observed velocity gradient along the Tuc III tails, and the debris in the Sutlej segments continues that trend. Segment 4 is predicted near $v_{\rm hel} \approx -100$ to $-85\,\mathrm{km\,s^{-1}}$, on the extension of the Tuc III gradient. Segment 2 is predicted at $\sim$$-100$ to $-70\,\mathrm{km\,s^{-1}}$, and segments 1 and 3 are offset to $\sim$$-90$ to $-50\,\mathrm{km\,s^{-1}}$.
\end{enumerate}

A Tuc III origin for Sutlej is therefore plausible and we make testable predictions to confirm its nature. Spectroscopy of Sutlej members would provide a more direct test for the coherence of the two stream populations.

\subsection*{acknowledgments}

The authors acknowledge support from the DiRAC Institute in the Department of Astronomy at the University of Washington. The DiRAC Institute is supported through generous gifts from the Charles and Lisa Simonyi Fund for Arts and Sciences, Janet and Lloyd Frink, and the Washington Research Foundation.

W.C. gratefully acknowledges support from a Gruber Science Fellowship at Yale University. This material is based upon work supported by the National Science Foundation Graduate Research Fellowship Program under Grant No. DGE2139841.

The DELVE project is partially supported by the NASA Fermi Guest Investigator Program Cycle 9 No.\ 91201 and by Fermilab LDRD project L2019-011. This material is based upon work supported by the National Science Foundation under Grant No.\ AST-2108168, AST-2108169, AST-2307126, and AST-2407526. This research award is partially funded by a generous gift of Charles Simonyi to the NSF Division of Astronomical Sciences. The award is made in recognition of significant contributions to Rubin Observatory's Legacy Survey of Space and Time. Any opinions, findings, and conclusions or recommendations expressed in this material are those of the author(s) and do not necessarily reflect the views of the National Science Foundation.

This project used data obtained with the Dark Energy Camera (DECam), which was constructed by the Dark Energy Survey (DES) collaboration. Funding for the DES Projects has been provided by the U.S. Department of Energy, the U.S. National Science Foundation, the Ministry of Science and Education of Spain, the Science and Technology Facilities Council of the United Kingdom, the Higher Education Funding Council for England, the National Center for Supercomputing Applications at the University of Illinois at Urbana-Champaign, the Kavli Institute of Cosmological Physics at the University of Chicago, the Center for Cosmology and Astro-Particle Physics at the Ohio State University, the Mitchell Institute for Fundamental Physics and Astronomy at Texas A\&M University, Financiadora de Estudos e Projetos, Funda\c{c}\~{a}o Carlos Chagas Filho de Amparo \`{a} Pesquisa do Estado do Rio de Janeiro, Conselho Nacional de Desenvolvimento Cient\'{\i}fico e Tecnol\'{o}gico and the Minist\'{e}rio da Ci\^{e}ncia, Tecnologia e Inova\c{c}\~{a}o, the Deutsche Forschungsgemeinschaft and the Collaborating Institutions in the Dark Energy Survey.

The Collaborating Institutions are Argonne National Laboratory, the University of California at Santa Cruz, the University of Cambridge, Centro de Investigaciones Energ\'{e}ticas, Medioambientales y Tecnol\'{o}gicas-Madrid, the University of Chicago, University College London, the DES-Brazil Consortium, the University of Edinburgh, the Eidgen\"{o}ssische Technische Hochschule (ETH) Z\"{u}rich, Fermi National Accelerator Laboratory, the University of Illinois at Urbana-Champaign, the Institut de Ci\`{e}ncies de l'Espai (IEEC/CSIC), the Institut de F\'{\i}sica d'Altes Energies, Lawrence Berkeley National Laboratory, the Ludwig-Maximilians Universit\"{a}t M\"{u}nchen and the associated Excellence Cluster Universe, the University of Michigan, NSF NOIRLab, the University of Nottingham, The Ohio State University, the University of Pennsylvania, the University of Portsmouth, SLAC National Accelerator Laboratory, Stanford University, the University of Sussex, Texas A\&M University, and the OzDES Membership Consortium.

Based in part on observations at NSF Cerro Tololo Inter-American Observatory at NSF NOIRLab (NOIRLab Prop.\ ID 2012B-0001, PI: J.~Frieman; NOIRLab Prop.\ ID 2019A-0305, PI: A.~Drlica-Wagner), which is managed by the Association of Universities for Research in Astronomy (AURA) under a cooperative agreement with the National Science Foundation.

The DES data management system is supported by the National Science Foundation under Grant Numbers AST-1138766 and AST-1536171. The DES participants from Spanish institutions are partially supported by MICINN under grants PID2021-123012, PID2021-128989, PID2022-141079, SEV-2016-0588, CEX2020-001058-M and CEX2020-001007-S, some of which include ERDF funds from the European Union. IFAE is partially funded by the CERCA program of the Generalitat de Catalunya. We acknowledge support from the Brazilian Instituto Nacional de Ci\^{e}ncia e Tecnologia (INCT) do e-Universo (CNPq grant 465376/2014-2).

This work has made use of data from the European Space Agency (ESA) mission \textit{Gaia} (\url{https://www.cosmos.esa.int/gaia}), processed by the \textit{Gaia} Data Processing and Analysis Consortium (DPAC, \url{https://www.cosmos.esa.int/web/gaia/dpac/consortium}). Funding for the DPAC has been provided by national institutions, in particular the institutions participating in the \textit{Gaia} Multilateral Agreement.

This work made use of the Claude large language model (Anthropic) to suggest language edits and draft portions of the manuscript text, and to assist in writing and debugging code for analysis and plotting. The authors take full responsibility for the content of this paper, and all scientific results and conclusions are their own.

\facilities{Blanco (DECam), \textit{Gaia}.}

\software{agama \citep{Vasiliev2019}, Astropy \citep{Astropy2013, Astropy2018, Astropy2022}, gala \citep{PriceWhelan2017}, HEALPix/healpy \citep{Gorski2005, Zonca2019}, Matplotlib \citep{Hunter2007}, nbody\_streams \citep{Arora2026}, NumPy \citep{Harris2020}, pandas \citep{McKinney2010, pandas2020}, ugali \citep{Bechtol2015}.}

\bibliographystyle{aasjournal}
\bibliography{refs}

\appendix
\renewcommand{\thefigure}{\Alph{section}\arabic{figure}}
\setcounter{figure}{0}

\section{Half-segment Hess diagrams}
\label{app:halves}

To look for distance gradients within each segment, we divide each segment at the midpoint of its two endpoints into a left (greater $L_{\mathrm{MS}}$) half and a right (lower $L_{\mathrm{MS}}$) half. Just as before, the on-stream sample is the stars within $\pm0.375^\circ$ in $B_{\rm MS}$ of the half-segment track, and the background comes from the same fixed fields BG1 and BG2, scaled by the ratio of unmasked areas measured on a HEALPix ($N_{\rm side} = 512$) grid. The Hess diagrams use the same 75 color $\times$ 80 magnitude bins and two-bin smoothing as Figure~\ref{fig:hess}. We fit the distance modulus of each half by eye, shifting the Sutlej isochrone in steps of 0.1\,mag. The left/right values are $(m-M) = 17.3/17.5$, $17.7/18.2$, $17.5/17.8$ and $16.8/17.5$ for segments 1--4, corresponding to 28.8/31.6, 34.7/43.7, 31.6/36.3 and 22.9/31.6\,kpc.

\begin{figure*}
\centering
\includegraphics[width=\textwidth]{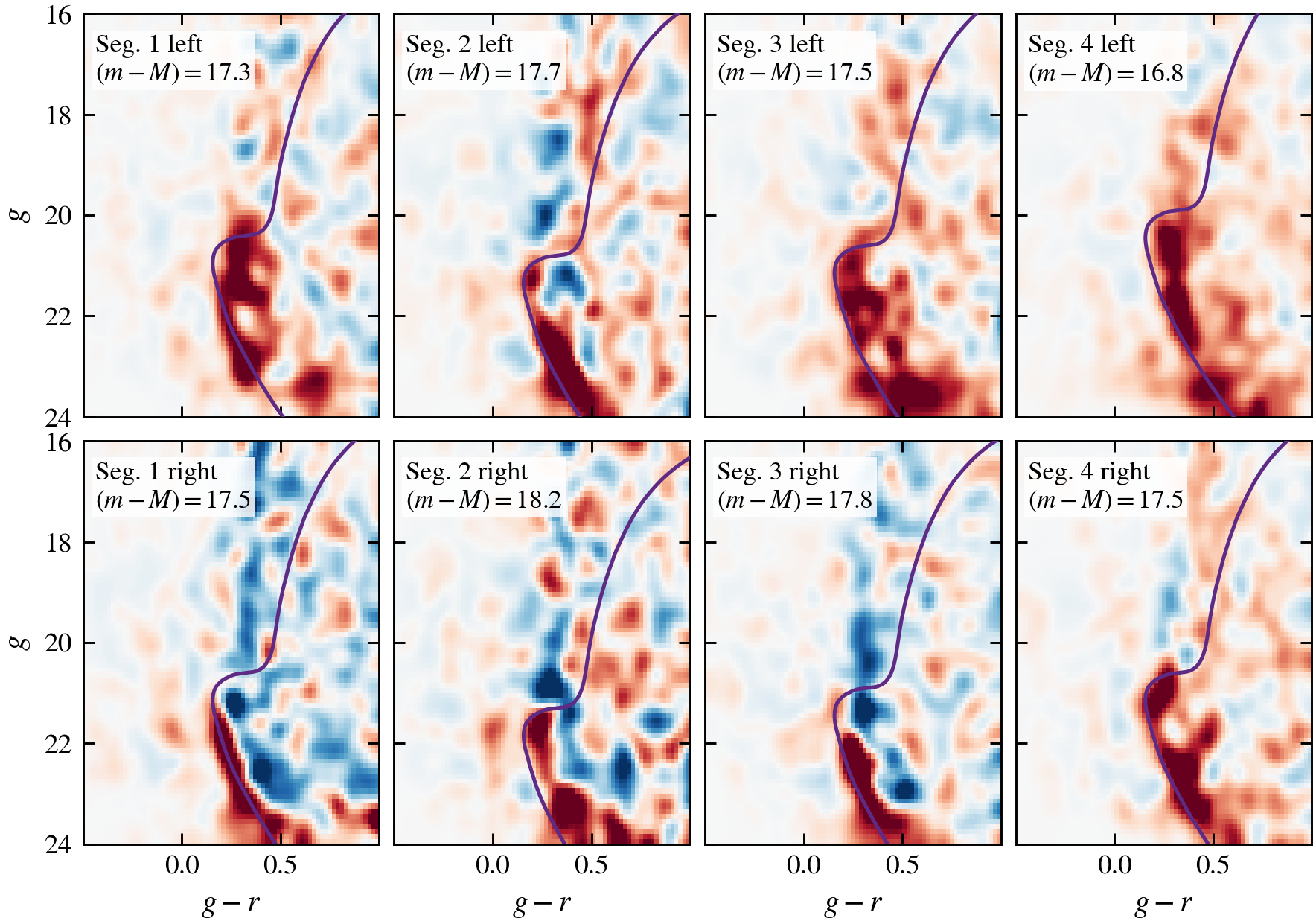}
\caption{Background-subtracted Hess diagrams of the half segments (red: excess over background; blue: deficit). Columns are segments 1--4. The top row shows the left halves and the bottom row the right halves, split at each segment's midpoint. The purple line is the Sutlej isochrone (\citealt{Marigo2017}; 10\,Gyr, $Z = 0.0003$) at each half's by-eye distance modulus, given in each panel. The binning, smoothing, and background fields are the same as in Figure~\ref{fig:hess}. These fits are the dotted lines in Figure~\ref{fig:dist}.}
\label{fig:halfhess}
\end{figure*}

\setcounter{figure}{0}
\section{Tuc III Simulation Details}
\label{app:tuc3}

Here we give the host potential and the priors on the orbit and progenitor used in Section~\ref{sec:tuc3}, and describe the two selection steps applied to the simulations: first keeping the realizations whose present-day remnant is consistent with Tuc III, and then identifying those whose debris resembles Sutlej.

\subsection{Host Potential and Orbit}

The bar adds only a non-axisymmetric component to the Milky Way potential, leaving the axisymmetric mass profile of the \citet{McMillan2017} model unchanged. We draw its pattern speed from a normal distribution with mean $41\,\mathrm{km\,s^{-1}\,kpc^{-1}}$ and standard deviation $3\,\mathrm{km\,s^{-1}\,kpc^{-1}}$ \citep{Sanders2019} and its present-day angle from a normal distribution with mean $27^\circ$ and standard deviation $2^\circ$ \citep{Wegg2013, BlandHawthorn2016}, although the bar has little effect on the debris and these choices do not affect our conclusions. The LMC orbit includes dynamical friction, and the simulations are run in the Milky Way frame with the corresponding reflex acceleration applied. We adopt the LMC distance and radial velocity from \citet{Shipp2021} and its \textit{Gaia} EDR3 proper motion \citep{GaiaLMC2021} with an uncertainty of $0.02\,\mathrm{mas\,yr^{-1}}$ per component, and draw its mass log-uniformly between $1 \times 10^{11}$ and $2.5 \times 10^{11}\,\mathrm{M}_\odot$.

For Tuc III, we sample the radial velocity and stream-track offset of the best-fit model of \citet[][Table A1]{Shipp2021} and the proper motion of \citet{Shipp2019} within their reported uncertainties, and draw the distance modulus uniformly between 16.7 and 17.3, a range deliberately broader than the measurement uncertainty on $(m-M) \approx 17.0$ \citep{DrlicaWagner2015, MutluPakdil2018}. Integrating the present-day position of Tuc III backward, every model passes the LMC about 75\,Myr ago at a closest approach of $\sim$10\,kpc, and the preceding pericenter, $\sim$0.5\,Gyr ago, reaches $\sim$0.3--1.5\,kpc, compared to $\sim$4\,kpc for the same orbits integrated without the LMC. The models that produce Sutlej-like debris (Appendix~\ref{app:sutlejsel}) are those with the deepest pericenters, with a median of 0.5\,kpc, and none of the models with a pericenter beyond 1\,kpc produce this debris. They also avoid the closest LMC encounters, with closest approaches of $\sim$11--14\,kpc and none within 10\,kpc, possibly because a closer passage scatters the debris too broadly to produce distinct overdensities in the segments. These orbits lie close to the observed one, concentrating at $16.9 \lesssim (m-M) \lesssim 17.1$, and with proper motions offset by only $\sim$0.5$\sigma$ from the measured value. The broad distance prior therefore does not drive the results, and the orbits that deposit debris in the Sutlej field are consistent with current measurements of Tuc III.

\subsection{Progenitor}

The progenitor is an NFW halo with an exponential truncation, $\exp[-(r/10r_\mathrm{s})^2]$, and a Plummer stellar component, sampled in equilibrium with \textsc{agama} \citep{Vasiliev2019} using 15{,}000 dark matter and 4{,}000 stellar particles. We assign the stellar component a total mass of $10^3\,\mathrm{M}_\odot$, following \citet{Shipp2021}. Because the stars are dynamically subdominant to the dark matter halo in all of our models, they act as tracers of the stripping, and their assigned mass does not affect the debris distribution. This mass is therefore a normalization for the simulations, not an estimate of the original stellar mass of Tuc III. We express stellar mass loss as the fraction of stellar particles that remain bound rather than as an absolute mass. We parameterize the halo by its mass within 37\,pc, the observed half-light radius of Tuc III \citep{MutluPakdil2018}, so that the prior can be compared directly with the dynamical upper limit of \citet{Simon2017}, and draw this mass log-uniformly between $10^4$ and $3 \times 10^5\,\mathrm{M}_\odot$, extending to a few times that limit. The scale radius is drawn log-uniformly between 50 and 500\,pc, and together these priors correspond to total halo masses of $\sim$$10^5$--$10^8\,\mathrm{M}_\odot$ at the start of the simulation, since these represent halos that may already have undergone substantial tidal stripping before the start of the simulations, rather than their masses at formation. The stellar Plummer radius is drawn uniformly between 20 and 50\,pc, bracketing the observed half-light radius, and the start time uniformly between 0.8 and 3\,Gyr ago, so that every model includes the most recent pericentric passage.

\subsection{Selecting Tuc III-like Remnants}

Most realizations do not end with anything resembling Tuc III, since the progenitor is either fully disrupted or leaves a diffuse cloud of debris rather than a bound remnant with a narrow tidal stream. The main constraint on the progenitor is therefore that Tuc III survives as a bound system, and we retain the realizations that end with a bound stellar remnant containing at least 1 percent of the stellar particles, a projected half-light radius within $2\sigma$ of the observed $37 \pm 9$\,pc \citep{MutluPakdil2018}, and a line-of-sight velocity dispersion below the upper limit of $1.5\,\mathrm{km\,s^{-1}}$ from \citet{Simon2017}. These criteria are intended to exclude models clearly inconsistent with the observed system rather than to match its properties, and the most heavily stripped remnants retain only tens of bound stellar particles, so their sizes and dispersions are too noisy to support tighter cuts. Of 499 completed realizations, 183 satisfy these criteria, and most also reproduce the radial velocity gradient measured along the Tuc III tails by \citet{Li2018}, which we do not use as a criterion.

\subsection{Selecting Sutlej-like Debris}
\label{app:sutlejsel}

\begin{figure*}
\centering
\includegraphics[width=\textwidth]{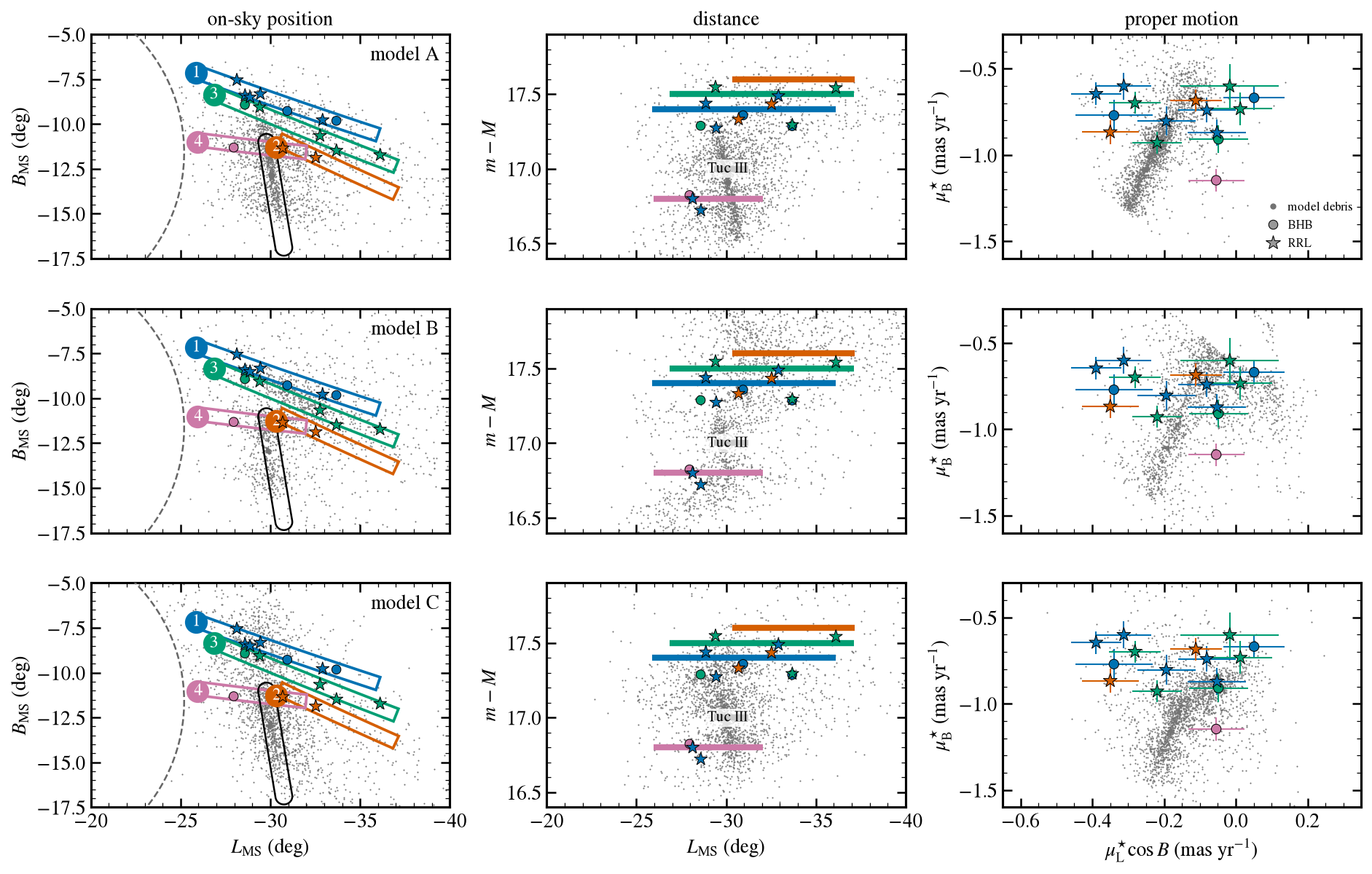}
\caption{Three examples of Tuc III models in which the debris occupies the spatial extent of Sutlej (rows), shown in on-sky position (left), distance modulus versus $L_{\rm MS}$ (center), and observed proper motion for model stars within the Sutlej segment boxes (right). Gray points are model stars, colored boxes and bars mark the four Sutlej segments and their fitted distance moduli, circles and stars mark the accepted BHB and RRL members respectively, the black outline marks the Tuc III stream, and the dashed arc marks the SMC mask. The models differ in the extent of the debris, but in each the debris grows more distant toward the trailing arm of Tuc III and occupies a narrow, elongated locus in proper motion along which most of the BHB and RRL stars fall.}
\label{fig:tuc3examples}
\end{figure*}

We next ask whether the debris of each retained model is concentrated in the Sutlej segments, rather than spread evenly across the field as a diffuse cloud or confined to a narrow band along the Tuc III orbit. For each segment, we compare the density of model stars in the segment box to that in the surrounding field, measured in strips of the same width on either side of the box starting $0.5^\circ$ from the segment track, and require all four segments to be overdense relative to the field outside their segment boxes. To compare at the observed distances, we count only model stars within $\pm 0.3$\,mag of each segment's measured distance modulus. Because debris along the unperturbed Tuc III orbit can produce an overdensity wherever it crosses a segment box, we evaluate the overdensity only along portions of the segments more than $1^\circ$ from the observed Tuc III stream. We apply this criterion to the simulations in the full potential with both the bar and the LMC. Of the 183 retained models, 182 completed reruns in all four host potentials, and 24 of these satisfy the criterion in the full potential, compared to 4 when the same criterion is applied to the reruns without the LMC. Figure~\ref{fig:tuc3examples} shows three of the 24 in on-sky position, distance, and proper motion.

%----------------------
\section{The Sutlej River}
\label{app:sutlej_origins}

The Sutlej stream was named by \citet{Nidever2024} after the Sutlej River, also spelled Satluj and known in Sanskrit as \'{S}atadru. Sutlej rises near Lake Rakshastal in Tibet and flows through Himachal Pradesh and Punjab, joining the Chenab and, through it, the Indus. It is the longest of the five rivers that give Punjab its name (from the Persian \textit{panj}, five, and \textit{\={a}b}, water). The river is central to the history, culture, and identity of the people of Punjab in both India and Pakistan \citep{Gilmartin2015, Kapuria2022}. Recent films reflect this, from the documentary \textit{The Mother, the Sons, and the Holy River} (2025), directed by Nirmal Bano and Natasha Javed, which tells the river's story through colonial rule, Partition, and the Indus Waters Treaty,\footnote{\url{https://www.imdb.com/title/tt39240118/}; see also \url{https://brownhistory.substack.com/p/the-sutlej-river-where-the-waters}} to \textit{Satluj} (2026), a film that follows Jaswant Singh Khalra's investigations into extrajudicial activities in Punjab in the early 1990s.\footnote{\url{https://varietyindia.com/diljit-dosanjh-on-satluj-it-took-me-a-week-to-move-on-after-this-film-exclusive/}} Naming a stellar stream for an earthly river inseparably links the two. We offer this note as a starting point for readers who wish to learn about the history of the river behind the name.

\end{document}